# Cultural Anthropology through the Lens of Wikipedia: Historical Leader Networks, Gender Bias, and News-based Sentiment

Peter A. Gloor, Joao Marcos, Patrick M. de Boer, Hauke Fuehres, Wei Lo, Keiichi Nemoto
pgloor@mit.edu
MIT Center for Collective Intelligence

**Abstract**

In this paper we study the differences in historical World View between Western and Eastern cultures, represented through the English, the Chinese, Japanese, and German Wikipedia. In particular, we analyze the historical networks of the World's leaders since the beginning of written history, comparing them in the different Wikipedias and assessing cultural chauvinism. We also identify the most influential female leaders of all times in the English, German, Spanish, and Portuguese Wikipedia. As an additional lens into the soul of a culture we compare top terms, sentiment, emotionality, and complexity of the English, Portuguese, Spanish, and German Wikinews.

## 1 Introduction

Over the last ten years the Web has become a mirror of the real world (Gloor et al. 2009). More recently, the Web has also begun to influence the real world: Societal events such as the Arab spring and the Chilean student unrest have drawn a large part of their impetus from the Internet and online social networks. In the meantime, Wikipedia has become one of the top ten Web sites[1], occasionally beating daily newspapers in the actuality of most recent news. Be it the resignation of German national soccer team captain Philipp Lahm, or the downing of Malaysian Airlines flight 17 in the Ukraine by a guided missile, the corresponding Wikipedia page is updated as soon as the actual event happened (Becker 2012. Futterer et al. 2013). In a long-term project at the MIT Center for Collective Intelligence, we are using Wikipedia to provide a mirror of today's historical understanding of World history by different cultures.

We are using Wikipedias in different languages as a window into the "soul" of different cultures, replacing anthropological fieldwork with statistical analysis of the treatment given by native speakers of a culture to different subjects in Wikipedia. One of the most popular categories in Wikipedia is the people pages, talking about the most important people of all ages. Wikipedians have put together "notability criteria" that clearly define if a person deserves inclusion into Wikipedia or not. In earlier work we have used the people pages in Wikipedia to identify the most influential thought leaders, idea givers, and academics through social network analysis techniques (Frick et al 2013), and have calculated maps of the most influential people in the different language Wikipedias (Kleeb et. al 2012).

[1] http://www.alexa.com/topsites

In this paper we introduce Wikihistory, a dynamic temporal map of the most influential people of all times in four different Wikipedias (English, German, Chinese, Japanese), and then look at the different distributions of gender in the English, Portuguese, Spanish, and German Wikipedia, resulting in a comparison among cultures on the subject of gender equality. In addition we also compare sentiment and emotionality in the English, Spanish, Portuguese, and German Wikinews, a hand-curated News page which is part of Wikipedia.

**1.1 Bias of Wikipedia**

For different parts of our analysis we rely on the English, German, Chinese, Japanese, Portuguese, and Spanish Wikipedias. While the German, Japanese, and Portuguese Wikipedias are more or less representative of their language spaces, things are more complex for the English, Chinese, and Spanish Wikipedias. Owing to the global dominance of the English language, the English Wikipedia is by far the largest, with 1,4 million monthly page edits in the US (with US population 308 million) and 486,000 page edits in the UK (with a UK population of 62 million). While the English Wikipedia could claim to reflect the dominant view of the World, it comes with a heavy bias towards worldview of the US and UK. This is in contrast to the Chinese Wikipedia, with 51,000 monthly edits in Hong Kong (with a population of 7 million), 46,000 monthly edits in Taiwan (with a population of 23 million) and 35,000 monthly edits in mainland China (with a population of 1.3 billion people). As Wikipedia is officially blocked in China, editing is done to a large part in Hong Kong and Taiwan, however, as we will see below, based on the importance of communist heroes in the Chinese Wikipedia, we suspect that Chinese censors are actively editing the Chinese Wikipedia.

The Spanish Wikipedia on the other hand draws on a widely distributed editor base in Spain, but also many contributors in the more populous Latin American countries such as Colombia and Mexico, and some very active editor communities in Chile and Argentina.
To a smaller extent English Wikipedians also exhibit some political bias, as they have a reputation of having a leftist liberal bent. In a study Greenstein and Zhu (2012) found that compared to the early days of Wikipedia, the liberal bias, while still there, has been reduced. They found that phrases like "civil rights" and "trade deficit" favored by Democrats are still more prominent than phrases like "economic growth" and "illegal immigration" favored by Republicans. We expect that the Japanese and German Wikipedias show a similar bias.

## 2 Methodology for Network Creation

Our goal was to create a social network of all notable people in humankind's history. Given our subjective view of the past and the fact that our collective memory forgets facts over time, we had to approximate this social network using a proxy: We only consider people that made it into Wikipedia, for whom we assume that they fulfill Wikipedia's notability criteria. In this social graph, people are considered vertices. For every underlying people page on Wikipedia, all references to other people are used to infer edges to other vertices. The amount of mentions is used to calculate the weight of each directed edge. As a second requirement, a link between two people can only exist if both of them were living at the same time. For the English Wikipedia, we therefore start with the 900,000 pages tagged as "people pages". All 900,000 people pages are dated, by extracting the dates of birth and of death of each individual. Using this information, 4900 networks through history, from 3000 BC to 1950 CE are calculated, as shown in figure 1.

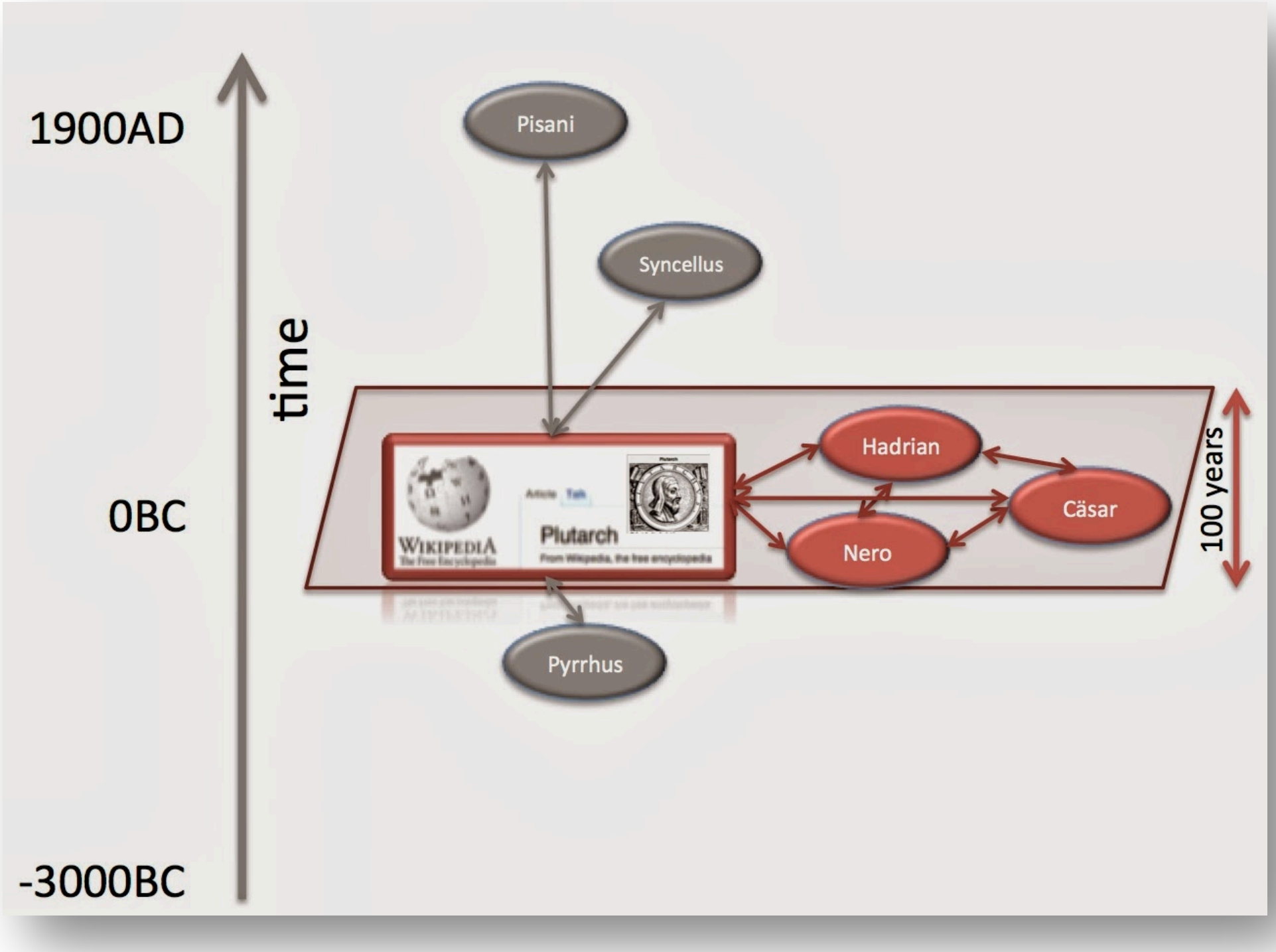

**Figure 1. Construction of social network of a historical personality**

For instance, in the graph shown in figure 1, from all the links to the page about Plutarch, only the links from and to Hadrian, Caesar, and Nero are kept, while the links to Pyrrhus, who died well before Plutarch was born, and the pages to medieval historian Syncellus and modern historian Pisani are ignored as well. Repeating this process leads to 4900 unique networks (less for the Chinese and Japanese Wikipedia, as their history does not go as far back). For each of these networks, the most central people in the graph are determined using the PageRank algorithm (Page et al. 1999). We consider central people as being highly influential to other people in the graph. To get a second selection criteria among all the influencers, their indegree, i.e. other people pages pointing back to them, is taken.

## 3 The Top 50 of All Times

Who are the most important people of all times? The answer to this question is quite different in the US, the UK, and German-speaking countries than it is in China and Japan. Looking at the top ten and top 50 people lists also confirms that most English language Wikipedia editors come from the US and the UK, while Chinese language editors come from Taiwan, Hong Kong, and mainland China.

### 3.1 English Wikipedia Analysis

The 50 most important people in the English Wikipedia from 3000BC of 1900AD are primarily politicians (26, kings and generals), followed by religious leaders (13), and poets and historians (11) (Table 1).

The disproportionally large role of the historians clearly stands out as they shape our view on the past. It seems that it pays to be a historian, to write one's own place in history. This is clearly shown by Sidney Lee, a relatively minor Victorian professor of English and history, who wrote 800 biographies and hereby secured his place in the annals.

Not only is a minor 19th century biographer under the top 10 influencers of all times (which is also an artifact of our collection method), but also classical historians like Polybius, Tacitus, and Plutarch get very high ranks. Treating biographers and historians well so they write positively about world leaders is of course no new insight: Roman emperor Vespasian was paying historians Tacitus, Suetonius, Josephus and Pliny the Elder, in return they speak suspiciously well about him, shaping his positive image in history. Caesar and Winston Churchill took this concept one step further, writing their history themselves. As today’s history is written in Wikipedia, the conclusion seems obvious: treat Wikipedians well!

The next three networking pictures illustrate the English Wikipedia Leadership networks through the ages, with snapshots taken at 600BC, at 0AD, and at 600AD. In correspondence with the availability of written records, in 600BC the Greek philosophers – documented in classical texts by Roman librarians – and Babylonian kings – documented in writings on the walls of historical buildings – are in the center, but there is also a noticeable cluster of Chinese leaders, as well as historical figures around Buddha.

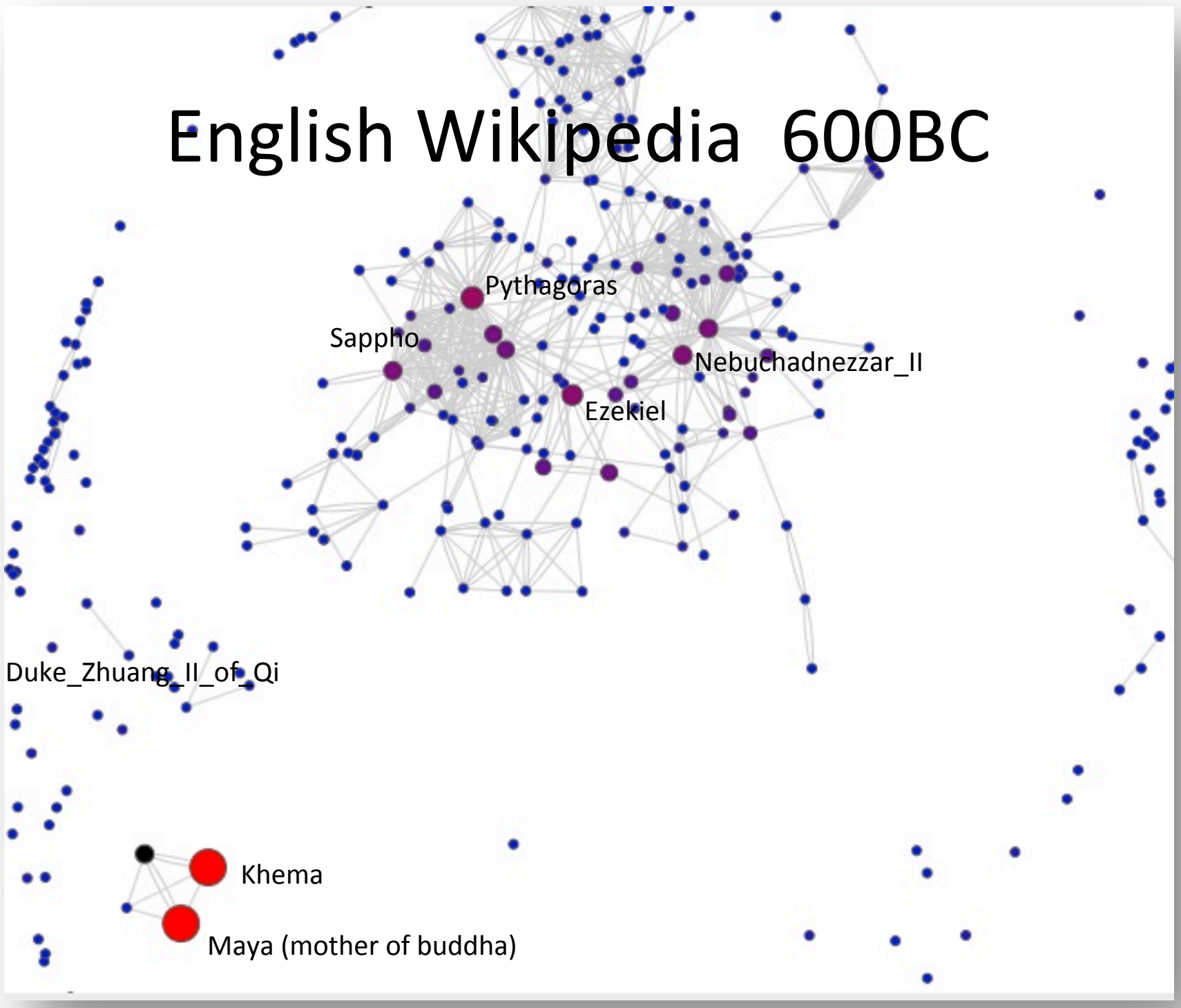

**Figure 2. Global leadership network of the English Wikipedia in 600BC**

Around 0AD, Roman emperors, consuls and writers are in the center, as well as biblical figures. Jesus has not been born yet, but his mother Mary has a prominent position in the biblical cluster.

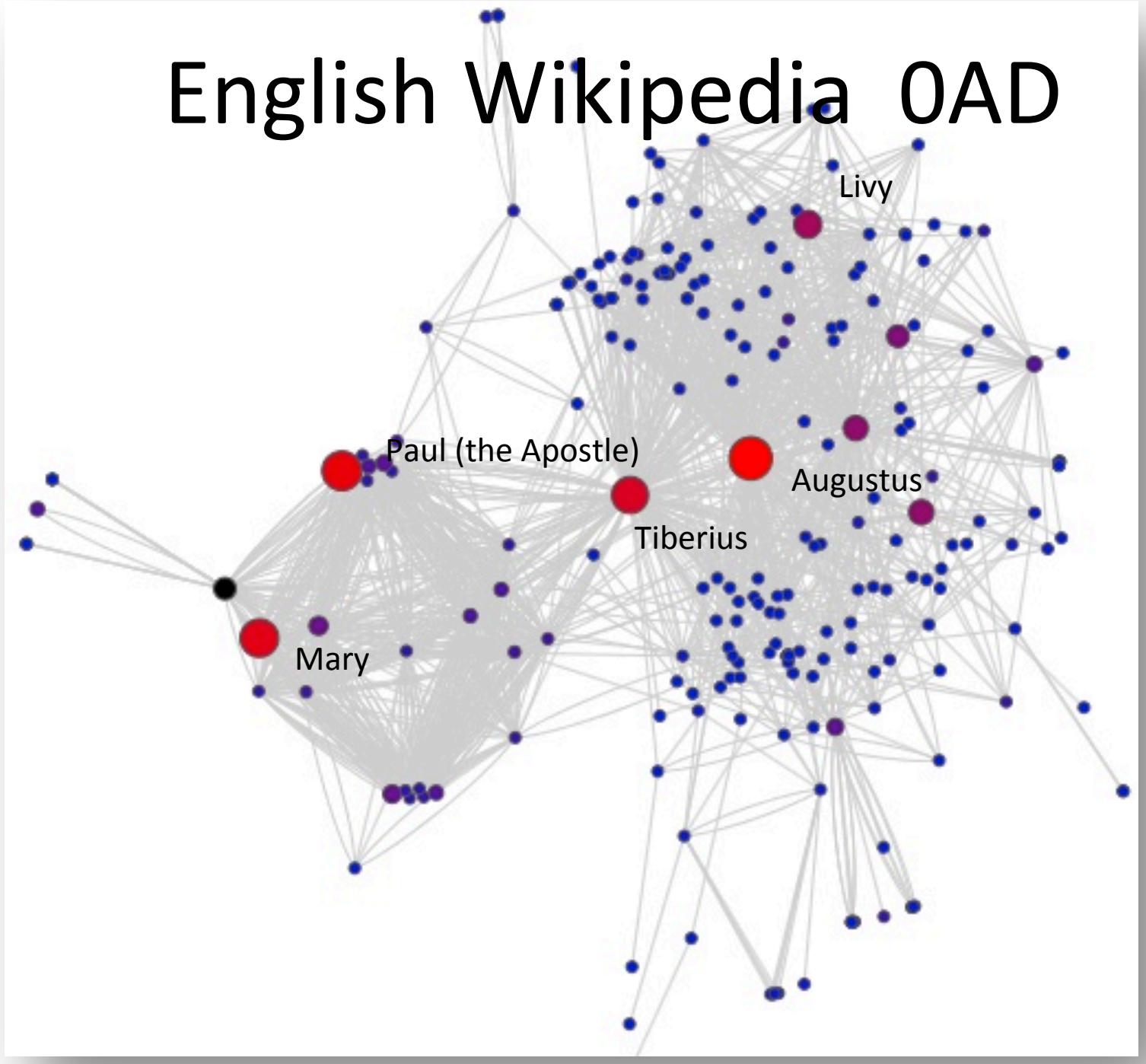

**Figure 3. Global leadership network of the English Wikipedia in 0AD**

Around 600AD, the Catholic Church with Popes, monks, and saints stands out. Greek orthodox church fathers around Babai the Great form a separate cluster. The Chinese emperors and generals form another cluster of almost similar size, representative for the subsequent blossoming of the Tang dynasty starting in 618, commonly seen as the high point in Chinese civilization.

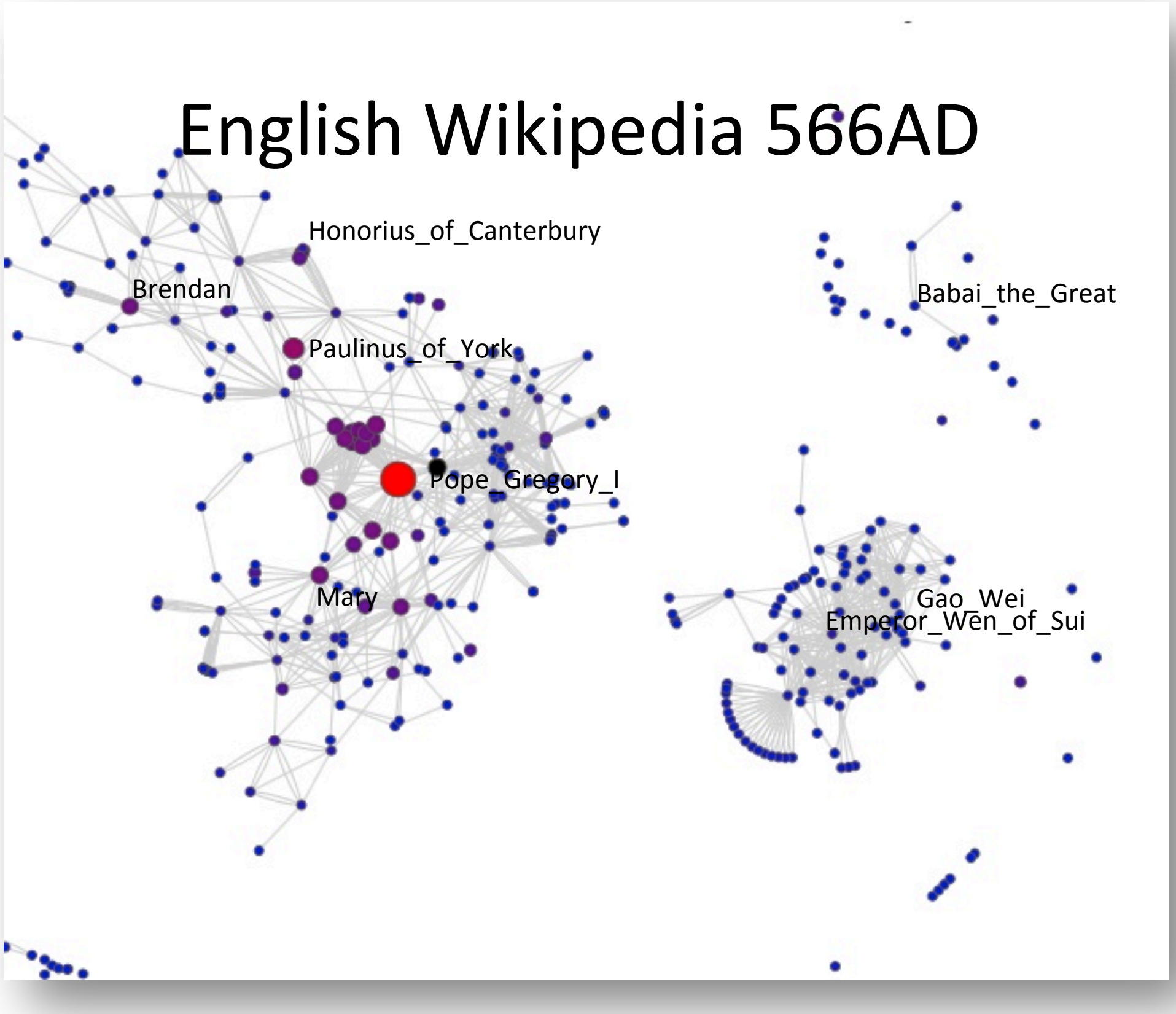

**Figure 4. Global leadership network of the English Wikipedia in 0AD**

| | English | Chinese | Japanese | German |
|---|---|---|---|---|
| 1 | George W. Bush | Mao Zedong | Ikuhiko Hata | Adolf Hitler |
| 2 | William Shakespeare | Yuan Shikai | Tokugawa Ieyasu | Johann Wolfgang von Goethe |
| 3 | Sidney Lee | Jay Chou | Toyotomi Hideyoshi | Aristoteles |
| 4 | Jesus | Oda Nobunaga | Adolf Hitler | Benedikt XVI. |
| 5 | Charles II of England | Tokugawa Ieyasu | Oda Nobunaga | Platon |
| 6 | Aristotle | Emperor Gaozong of Tang | Hirohito | Martin Luther |
| 7 | Napoleon | Cao Cao | Tokugawa Hidetada | Otto von Bismarck |
| 8 | Muhammad | Kangxi Emperor | Tokugawa Iemitsu | Johannes Paul II. |
| 9 | Charlemagne | Emperor Huizong of Song | Chiang Kai-shek | Johann Heinrich Zedler |
| 10 | Plutarch | Yongle Emperor | Tokugawa Ienari | Johann Sebastian Bach |
| 11 | Julius Caesar | Kangxi Emperor | Emperor Meiji | Wilhelm II. (Deutsches Reich) |
| 12 | William III of England | Hongwu Emperor | Tokugawa Tsunayoshi | Karl V. (HRR) |
| 13 | Homer | Jiajing Emperor | Tokugawa Yoshimune | Wolfgang Amadeus Mozart |
| 14 | Bede | Koxinga | Emperor Go-Daigo | Napoleon Bonaparte |
| 15 | Athanasius of Alexandria | Wang Shichong | Minamoto no Yoritomo | Richard Wagner |
| 16 | Dante Alighieri | Emperor Daizong of Tang | Emperor Go-Shirakawa | Georg Wilhelm Friedrich Hegel |
| 17 | Gautama Buddha | Emperor Xuanzong of Tang | Ashikaga Takauji | Marcus Tullius Cicero |
| 18 | Tiberius | Zhengtong Emperor | Ashikaga Yoshimitsu | Pius XII. |
| 19 | Cyril of Alexandria | Emperor Xizong of Tang | Emperor Go-Toba | Ludwig van Beethoven |
| 20 | Bernard of Clairvaux | Qianlong Emperor | Emperor Saga | Karl Marx |
| 21 | Moses | Dong Zhuo | Ashikaga Yoshimasa | Friedrich Schiller |
| 22 | Tacitus | Yuwen Tai | Ashikaga Yoshinori | Joseph Goebbels |
| 23 | Edward III of England | Qianlong Emperor | Emperor Toba | Abraham Lincoln |
| 24 | Justinian I | Huan Xuan | Emperor Kanmu | Heinrich Himmler |
| 25 | David | Emperor Huizong of Song | Emperor NinmyÅ | Denis Diderot |
| 26 | Ashoka | Emperor Xuanzong of Tang | Emperor Daigo | Benito Mussolini |
| 27 | Origen | Emperor Renzong of Song | Emperor Junna | Thomas Jefferson |
| 28 | Septimius Severus | Temür Khan, Emperor Chengzong of Yuan | Emperor Seiwa | Immanuel Kant |
| 29 | Polybius | Emperor Taizong of Song | Emperor Wu of Liu Song | Friedrich II. (HRR) |
| 30 | Confucius | Fu Jian | Emperor Montoku | Heinrich Heine |
| 31 | Alexander Severus | Sima Ying | Emperor Uda | Bertolt Brecht |
| 32 | Patriarch Eutychius of Alexandria | Emperor Taizong of Song | Fujiwara no Michinaga | Andrew Jackson |
| 33 | Tutankhamun | Wang Mang | Emperor Wen of Liu Song | Gerhart Hauptmann |
| 34 | Akhenaten | Emperor Shunzong of Tang | Cao Cao | Franziskus (Papst) |
| 35 | Ramesses II | Shi Jingtang | Emperor Xiaowu of Liu Song | Ludwig XIV. |
| 36 | Pope Benjamin I of Alexandria | Wang Yangming | Emperor Xiaowen of Northern Wei | Franz Liszt |
| 37 | Teti | Hong Xiuquan | Emperor Ming of Liu Song | Konrad Adenauer |
| 38 | Amenemhat II | Napoleon III | Emperor Wu of Southern Qi | Plutarch |
| 39 | Pepi II Neferkare | Emperor Xiaowu of Liu Song | Emperor Wu of Liang | George Washington |
| 40 | Merneith | Sima Shi | Sun Quan | Pablo Picasso |
| 41 | Terence | Möngke Khan | Augustus | Friedrich Nietzsche |
| 42 | Cato the Elder | Chang Yuchun | Pompey | Albert Einstein |
| 43 | Charles Martel | Emperor Ming of Han | Tiberius | Barack Obama |
| 44 | Gilgamesh | Tuoba Huang | Mark Antony | Augustus |
| 45 | Deborah | Yuwen Huaji | Alexander the Great | Alexander von Humboldt |
| 46 | Lugalbanda | Mu Ying | Plato | Josef Stalin |
| 47 | Kubaba | Emperor Wen of Liu Song | Cicero | Heinrich IV. (HRR) |
| 48 | Fu Xi | Zhu Xi | Emperor Wu of Han | Benjamin Franklin |
| 49 | Henry I of England | Confucius | Ptolemy I Soter | Angela Merkel |
| 50 | Petrarch | Emperor Ming of Southern Qi | Octavia the Younger | Friedrich Engels |

**Table 1: Top 50 most important people of all times in 4 Wikipedias (most important at the top) Red denotes politicians, black religious leaders, blue scientists and artists**

## 3.2 Chinese Wikipedia Analysis

Eminent leaders in the Chinese Wikipedia show a markedly different picture. Other than in the English Wikipedia, the top 50 leaders of all times are mostly emperors and generals. While in the English Wikipedia only 26 out of 50 people fell into that category, in the Chinese Wikipedia there are only 4 people who are not political and military leaders, among them three Confucian scholars including Confucius himself, plus one of the most popular

Chinese pop musicians (who is mostly unknown outside Chinese speaking countries). As tables 1 and 2 illustrate, in China and Japan only famous warriors and politicians had a chance to make it into the top ten and top fifty – the East seems far less religious than the West in their view on the past – while the Western Wikipedias are more balanced with half of the top ten as well as the top fifty of all times being religious leaders, artists or scientists. Historians play a special role. Both Sidney Lee and Ikuhiko Hata, a 19 century Japanese military biographer, owe their prominent position to their prolific biography writing, as they get many backlinks from the references on the pages of contemporary politicians they wrote about.

Note that due to the smaller size of the Chinese Wikipedia, the indegrees of the top 50 Chinese leaders are much lower than for the English Wikipedia leaders. This means that in the lower ranks of the top 50 a few more incoming links can have a large effect on the ranking of a person. Also Chinese leaders do not go back as far in history as the Egyptian pharaohs and the ancient kings of the middle east, where a combination of year-ring dating and listing of dynasties allows to date ancient leaders back to 3000BC. The oldest entries in both Wikipedias which are Chinese are from 800BC. Mythical earlier emperors like the Yellow Emperor who reigned from 2698 to 2598 BC are not included in the dead people category, as historians are not certain if he ever really existed as a single real person.

The Chinese Wikipedia leaders have far fewer links to people outside China than leaders on the non-Asian Wikipedias we have looked at. One could therefore see them as more inward-looking, i.e. China focused. Napoleon III (not Napoleon I) and Tokugawa Ieyasu are the only non-Chinese/Japanese among the top 50 leaders. The English Wikipedia is far more diverse, including Confucius, Buddha, and Fu Xi besides military leaders from ancient Egypt and Sumeria together with the leaders of the Western World.

A big restriction of the Chinese Wikipedia is the censorship executed by the Chinese government, which might influence the scant inclusion of Western politicians, scientists, and religious leaders.

### 3.3 Japanese Wikipedia Analysis

Other than the Chinese Wikipedia, the Japanese Wikipedia is not subject of any censorship, and can thus rightfully claim to represent the opinion and collective intelligence of the Japanese population. The top person article by in-degree, just like the article about the historian Sidney Lee in the English Wikipedia, is about a military historian who wrote about all the politicians, mostly generals, referenced in the Japanese Wikipedia. Amazingly, the fourth most central article in the Japanese Wikipedia is about Adolf Hitler.

### 3.4 Cultural Chauvinism

There are striking differences in the number of out-group leaders, i.e. leaders not part of the language sphere of a particular Wikipedia, included among the top 50. While the English Wikipedia includes 80% non-English leaders among the top 50, only two non-Chinese made it into the top 50 of the Chinese Wikipedia: Napoleon III and Tokugawa Ieyasu. The Japanese Wikipedia is slightly more balanced, with almost 40 percent non-Japanese leaders, half of them Chinese Emperors, and people like Adolf Hitler, Plato, Cicero, and Augustus.

| | English | Chinese | Japanese | German |
|---|---|---|---|---|
| Politicians | 26 | 46 | 47 | 23 |
| Religious Leaders | 11 | 1 | 0 | 5 |
| Artists/ Scientists | 13 | 3 | 3 | 22 |
| Cultural In-group | 10 | 48 | 31 | 31 |

**Table 2: Distribution of different people categories in 4 Wikipedias among the top 50 people of all times**

A part of that effect might also come from the different sizes and ethnic profiles of the editors of the different Wikipedias, as the English Wikipedia probably includes articles written about Chinese leaders written by Chinese and Japanese living abroad, while the articles in the Chinese and Japanese Wikipedia are most likely written by ethnic Chinese and Japanese.

## 4 Differences in Gender Equality

In a subsequent analysis, we studied the percentage of female leaders in the different cultures. Although, in theory, Wikipedia people pages are tagged by gender, many of them don't include this mapping. We therefore created a heuristic algorithm that collects keywords in different languages and analyzes the frequency of those words on each page. It counts, for example “she”, “he”, ”herself”, etc., for the English Wikipedia. In the Portuguese Wikipedia we used words such as “ela”, “ele”, etc. The same approach was used for the German and Spanish Wikipedias. To verify the accuracy of this method we compared the results with a subset of people pages that include the gender tag. We found that this approach got the gender right in more than 90% of the people pages in those four languages.

### 4.1 Top 50 Gender Analysis

While many societies have made great progress towards gender equality over the last 100 years, we wanted to measure the pace of this progress. We tracked the evolution of the share of women among the people represented in the different Wikipedias. We found that the 50 most important women in the Wikipedias from 2000BC to 2000AD are almost exclusively in two categories, namely politicians (mostly queens) and artists (writers, actresses and singers for example).

| | German | English | Spanish | Portuguese |
|---|---|---|---|---|
| politicians | 10 | 14 | 12 | 11 |
| artists | 38 | 36 | 35 | 34 |
| %-politicians | 20.83% | 28.00% | 25.53% | 24.44% |

**Table 3. Split between politicians and artists among the top 50 most influential women of all times in different Wikipedia s**

Table 4 lists all the names in the four Wikipedias. The first group consists of politicians and queens. In this group the most prominent member is Queen Elisabeth II, showing that people's fascination with the British Royals is unbroken. In our metrics, she consistently leads the female leadership ranking. Also, in the German and Portuguese Wikipedias we find German chancellor Angela Merkel and Brazilian president Dilma Rusself among the top 5 positions in their corresponding countries, illustrating that in the last decades' women have been accumulating more influence in global politics. Another interesting result is the position of American politician Hillary Clinton. She is present among the most influential top 50 women in all Wikipedias but the English. We can speculate that maybe she has more influence overseas than inside her own country due to her position as foreign minister. We also have to note that this analysis was completed 2 years before Hillary announced her aspirations for the US 2016 Presidential race.

In the art categories we find a more diversified distribution between writers, singers and actresses. It also seems that the United States exerts a dominating influence in this area in the rest of the world. Many actresses that worked or work in the US show business, such as Marilyn Monroe, Judy Garland, or Barbra Streisand, have a high PageRank in the different Wikipedias.

| | German | English | Spanish | Portuguese |
|---|---|---|---|---|
| 1 | Elisabeth II. | Elizabeth II | Michelle Bachelet | Maria II de Portugal |
| 2 | Angela Merkel | Queen Victoria | Marilyn Monroe | Madonna |
| 3 | Maria Theresia | Joan Baez | Isabel I de Inglaterra | Isabel II do Reino Unido |
| 4 | Katie Fforde | Simone Signoret | María Teresa I de Austria | Dilma Rousseff |
| 5 | Marlene Dietrich | Benazir Bhutto | Mary Pickford | Maria I de Portugal |
| 6 | Rosamunde Pilcher | Eudora Welty | Katharine Hepburn | Vitória do Reino Unido |
| 7 | Margaret Thatcher | Gertrude B. Elion | Elizabeth Taylor | Judy Garland |
| 8 | Hannah Arendt | Betty White | Hillary Clinton | Cher |
| 9 | Greta Garbo | Annette Bening | Margaret Thatcher | Fernanda Montenegro |
| 10 | Elfriede Jelinek | Eunice Kennedy Shriver | Greta Garbo | Barbra Streisand |
| 11 | Sigrid Roth | Ruth Bader Ginsburg | María Estela Martínez de Perón | Meryl Streep |
| 12 | Maria (Mutter Jesu) | Susan Hayward | Vivian Malone Jones | Marilyn Monroe |
| 13 | Thea Leitner | Mary J. Blige | Eva Perón | Gal Costa |
| 14 | Ella Fitzgerald | Margaret Beckett | Madonna | Sophia Loren |
| 15 | Maria Stuart | Barbara Jordan | Gabriela Mistral | Britney Spears |
| 16 | Marilyn Monroe | Zora Neale Hurston | Barbra Streisand | Marta Suplicy |
| 17 | Jane Fonda | Rosalynn Carter | Britney Spears | Christina Aguilera |
| 18 | Bette Davis | Sarah Vaughan | Jane Fonda | Roberta Flack |
| 19 | Elizabeth Taylor | Heidi Klum | Marlene Dietrich | Maggie Smith |
| 20 | Billie Holiday | Helen Clark | Juana de Arco | Rihanna |
| 21 | Nelly Sachs | Donna de Varona | Whitney Houston | Helen Mirren |
| 22 | Liza Minnelli | Julie Walters | Virginia Woolf | Elis Regina |
| 23 | Diana Ross | Patti Smith | Christina Aguilera | Kate Winslet |
| 24 | Katharine Hepburn | Harriet Tubman | Mariah Carey | Maria Teresa da Áustria |
| 25 | Beatrix (Niederlande) | Michelle Pfeiffer | George Sand | Angelina Jolie |
| 26 | Joan Crawford | Allison Janney | Meryl Streep | Lauren Bacall |
| 27 | Katarina Witt | Carole King | Aretha Franklin | Whitney Houston |
| 28 | Shirley MacLaine | Alice Munro | Ella Fitzgerald | Judi Dench |
| 29 | Elisabeth I. | Faye Dunaway | Shakira | Marília Pêra |
| 30 | Romy Schneider | Willa Cather | Josefina de Beauharnais | Janet Jackson |
| 31 | Claudette Colbert | Margaret Sanger | Ingrid Bergman | Isabel I de Castela |
| 32 | Toni Morrison | Kirstie Alley | Judy Garland | Nigar Jamal |
| 33 | Ingrid Bergman | Joanna of Castile | Liza Minnelli | Jennifer Lawrence |
| 34 | Barbra Streisand | Sheryl Crow | Simone de Beauvoir | Cate Blanchett |
| 35 | Whoopi Goldberg | Sally Ride | María Josefa de Austria | Mary Pickford |
| 36 | Sigrid Undset | Jennifer Hudson | Ava Gardner | Maria Antonieta |
| 37 | George Sand | Serena Williams | Billie Holiday | Penélope Cruz |
| 38 | Nadine Gordimer | Donna Shalala | Samuel Johnson | Sissy Spacek |
| 39 | Gabriela Mistral | Shirley Chisholm | Lady Gaga | Whoopi Goldberg |
| 40 | Doris Lessing | Olivia Newton-John | Angela Merkel | Hillary Clinton |
| 41 | Pearl S. Buck | Kristin Scott Thomas | Nicole Kidman | Jennifer Lopez |
| 42 | Marie Antoinette | Margaret, Maid of Norway | Maria Callas | Ingrid Bergman |
| 43 | Helen Hayes | Candice Bergen | Vanessa Redgrave | Barbara Stanwyck |
| 44 | Meryl Streep | Wilma Rudolph | Joan Crawford | Isabel I de Inglaterra |
| 45 | Hillary Clinton | Martha Griffiths | Mercedes Sosa | Jodie Foster |
| 46 | Judy Garland | Antonina Houbraken | Janet Jackson | Shirley MacLaine |
| 47 | Alice Munro | Claire Danes | Rita Hayworth | Francisca de Bragança |
| 48 | Britney Spears | Marion Cotillard | Julia Roberts | Sandra Bullock |
| 49 | Goldie Hawn | Sappho | Tina Turner | Januária Maria de Bragança |
| 50 | Isabella I. (Kastilien) | Alice Hamilton | Rihanna | Lady Gaga |

**Table 4. Top most important female leaders of all times in four Wikipedias**

### 4.2 Longitudinal Gender Analysis

By analyzing the percentage of women over the last century, we can see how the fraction of women among the World's leaders is increasing over time (Figure 2). Our data shows the Portuguese Wikipedia having the highest percentage on women among the Wikipedias we analyzed, followed by the Spanish and English.

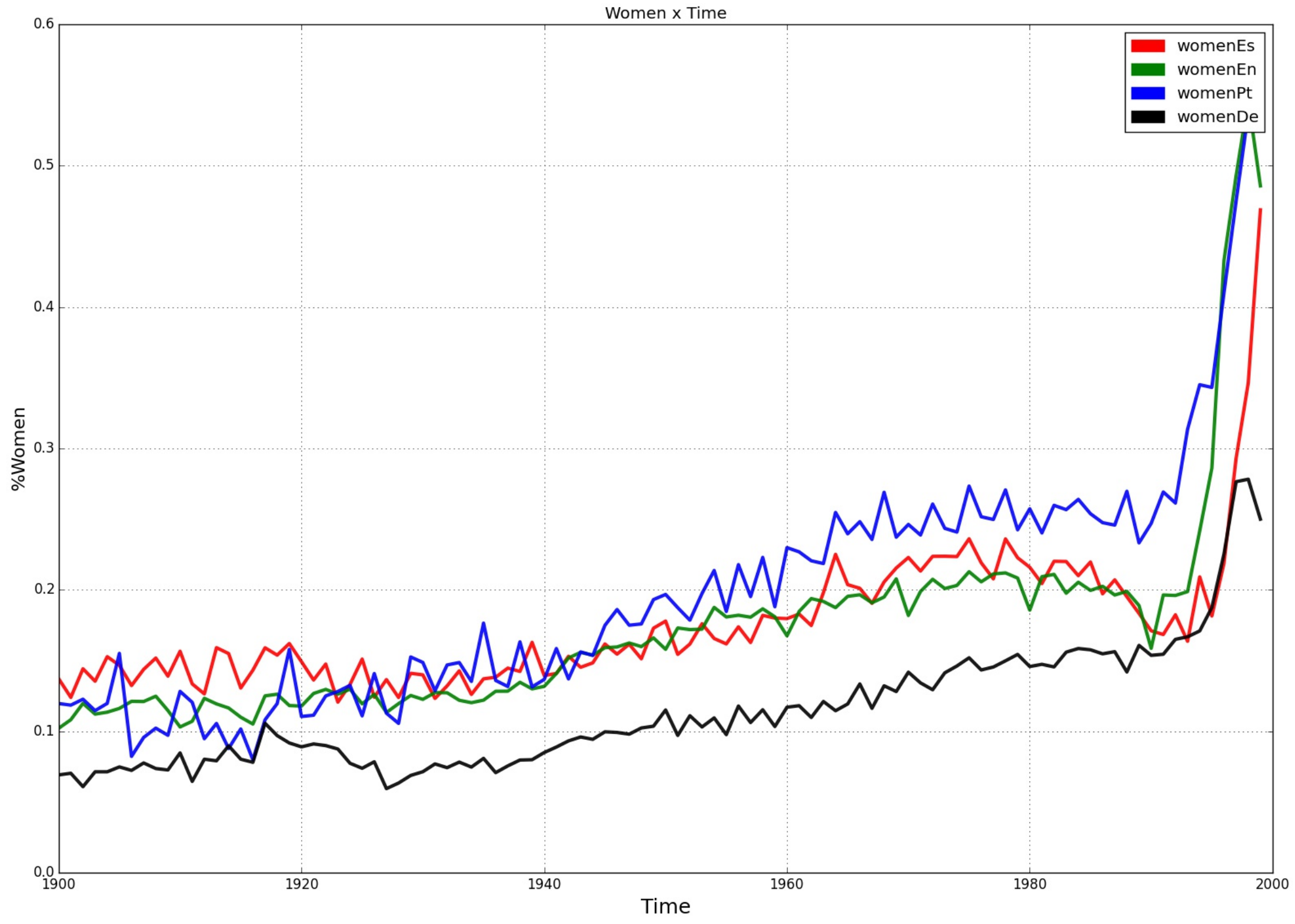

**Figure 5. Percentage of women among Wikihistory leaders from 1900 to 2000**

One fact that may have influenced this result is the number of total people pages in the Portuguese Wikipedia (93,000), which is significantly lower than the English Wikipedia having ten times more people pages, translating to a lower sample size. The graph shows a linear increase in the percentage of women until 1995, when the percentage really explodes to parity between women and men in the English, Spanish, and Portuguese Wikipedias. This explosion, however, might also be an artifact of the smaller number of people pages for the last 20 years, as there are very few people, other than Justin Bieber or Malala Yousafzai (the youngest ever peace Nobel-Prize winner), that fulfill the notability criteria of Wikipedia at such a young age. Nevertheless, it seems that in this smaller sample of young high-achievers we attain gender equality.

## 5 Comparing Different Cultures through Wikinews

We complement our historical analysis of Wikpedia with an analysis of the treatment of late breaking news by Wikipedians in Wikinews, a hand-curated news page of whatever Wikipedians consider most newsworthy on any given day. In particular, we use Wikinews to analyze sentiment and emotion in different cultures. Wikinews is part of Wikipedia. It is a free and open news web site where users create the content based on digesting news from sources such as Reuters, Bloomberg, and CNN. The main focus in this project is to analyze what kinds of news are more relevant in each culture and the sentiment related to those news. For example, we want to investigate if some cultures focus more on bad or good news. Or, whether some topics are considered more newsworthy in some cultures than in others.

### 5.1 Data Collection

We collected the news using the Wikinews pages in the English, German, Spanish and Portuguese. In those pages, we can find the most relevant events of each year through the eyes of the Wikipedians. In particular, whenever they consider a particular topic newsworthy, they will add a short sentence on the Wikinews page as an anchor, and create a new Wikipedia page for the entire event. For each language, we collected all the links to the pages in Wikipedia and connected those links with each other. This approach creates four distinct networks, one for each language (see Figures 6 to 9 bellow). As we can see, some Wikipedias have a more complex and connected news network. This is not surprising because for example the English Wikipedia has more content and is more updated than the Portuguese. The total number of articles is also quite different between the different Wikipedias.

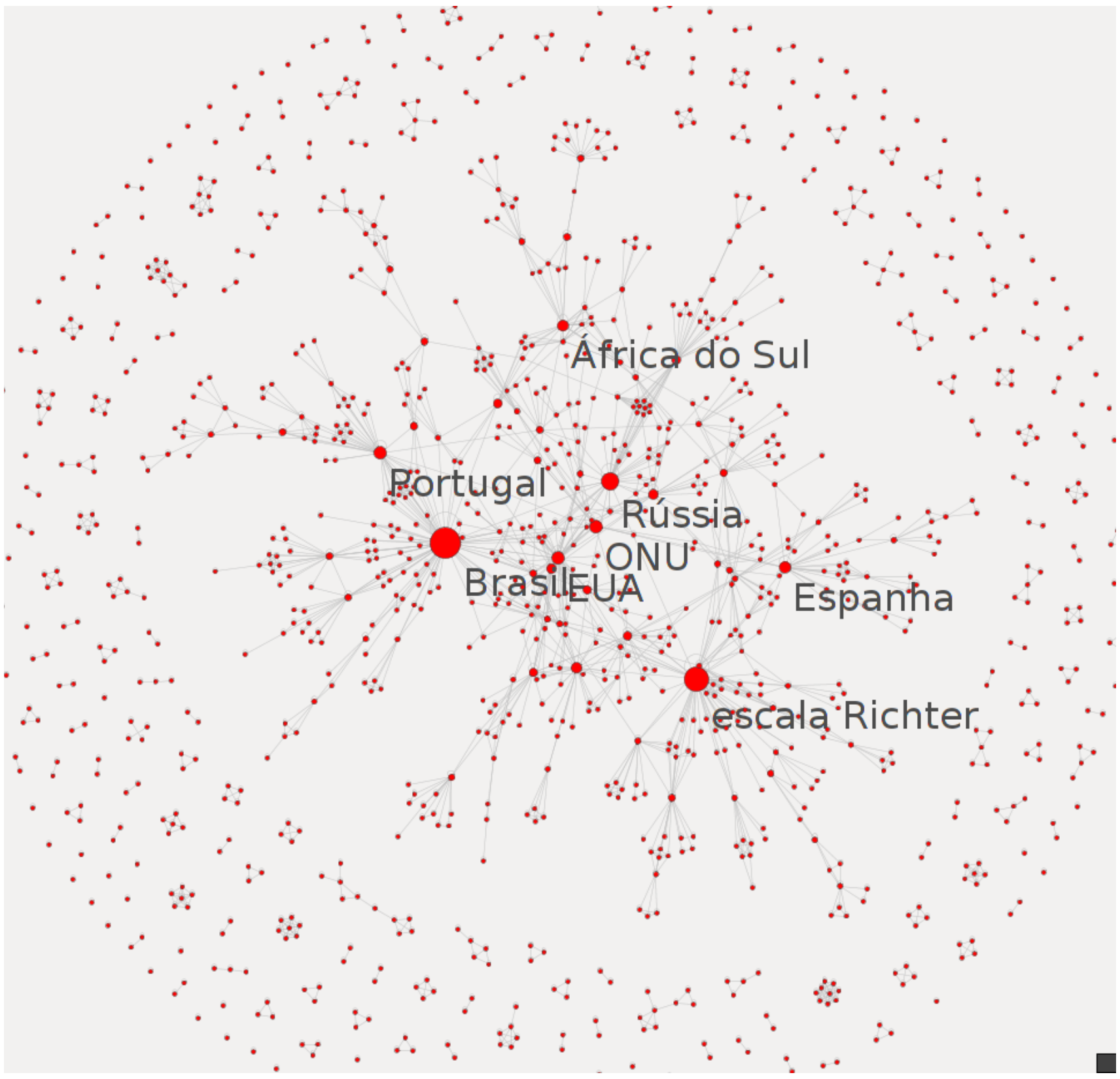

**Figure 6. Portuguese Wikinews network**

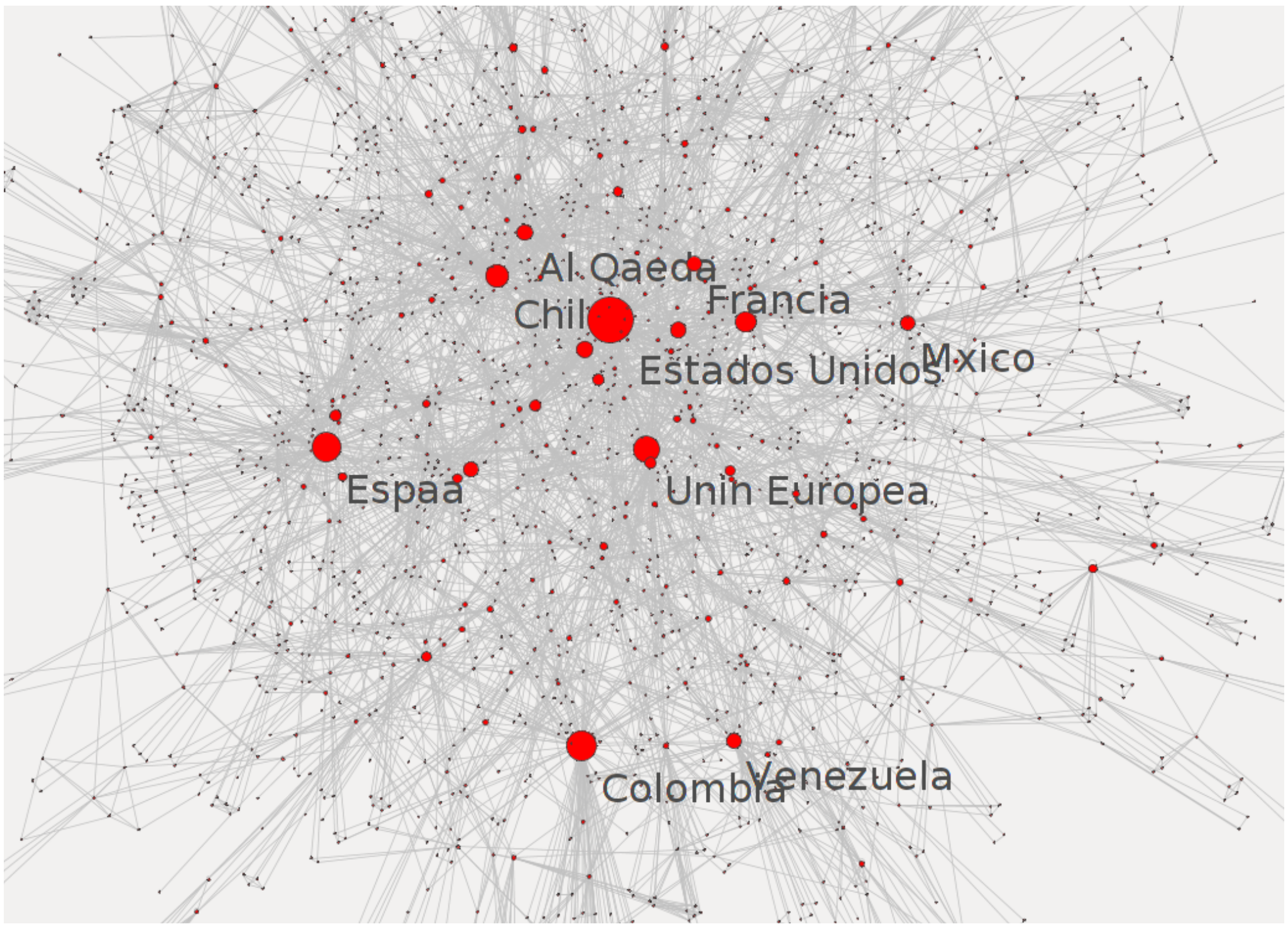

**Figure 7. Spanish Wikinews network**

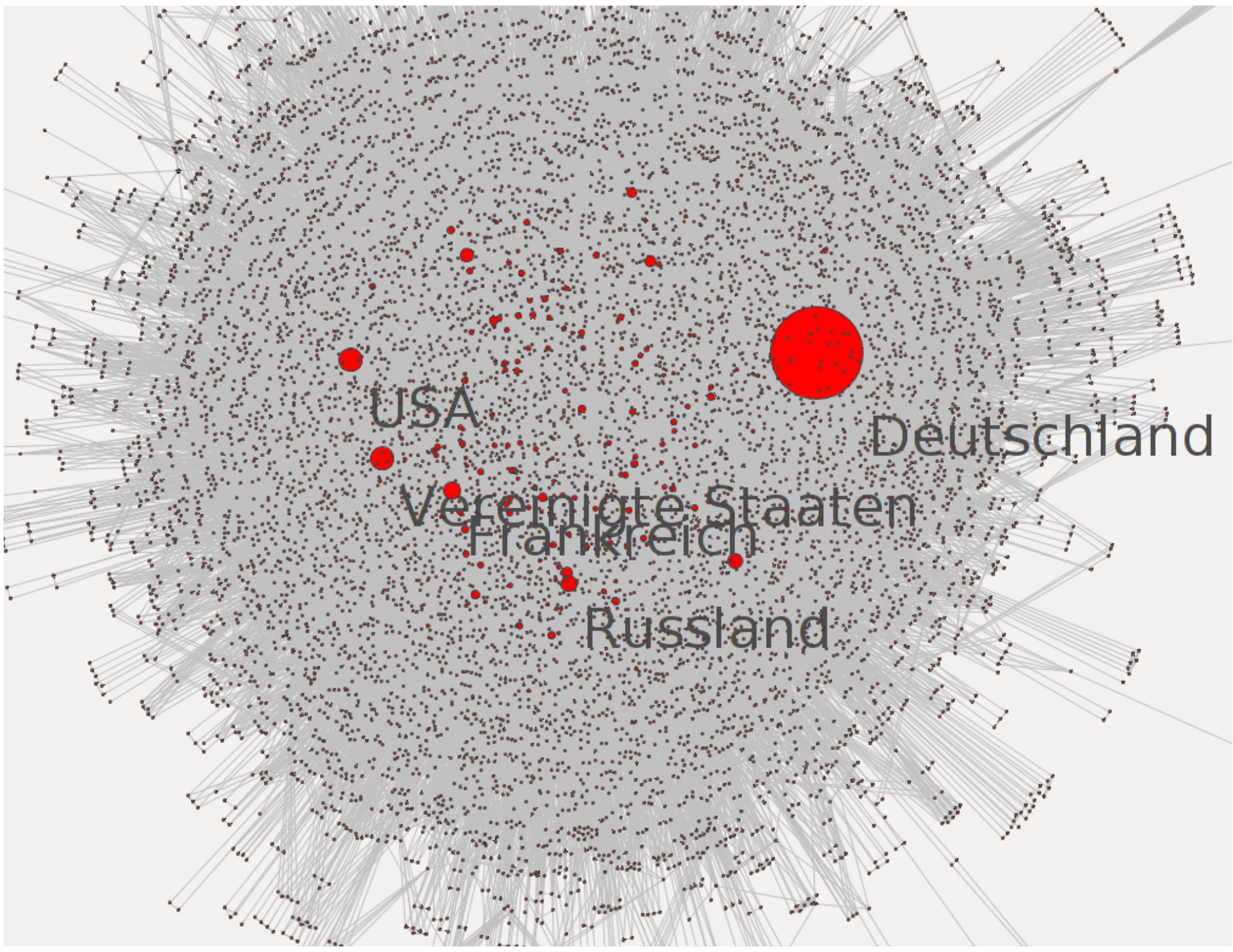

**Figure 8. German Wikinews network**

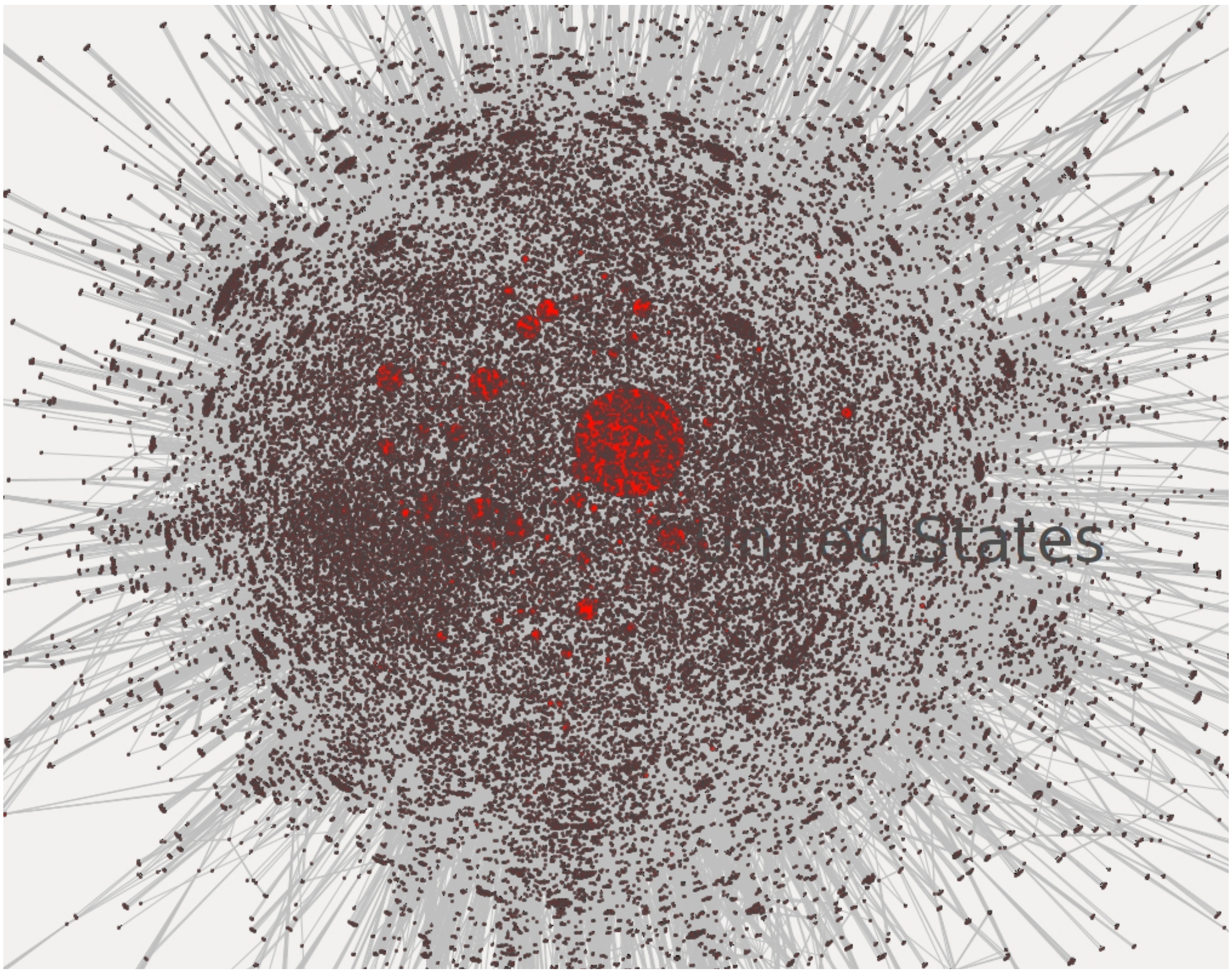

**Figure 9. English Wikinews network**

As we can see in the network pictures in Figures 6 to 9, the most central nodes are usually countries. News are linked to location pages related to where an event happened as opposed to one main subject like war, company names, or people. Unsurprisingly, we observe a heavy location bias, in that intra-cultural news get much more detailed treatment. This means that Wikipedia editors of one language upload news that happened in the country with the same language more frequently than outside news (Figure 10). For the Spanish Wikipedia, due to the many countries sharing the same language, we observe a more equal distribution of topics by betweenness centrality. As many different countries are appearing on the Spanish Wikinews with similar centrality, this illustrates that people from many of those countries seem to be contributing. The "popularity" of one country in other cultures and languages can show how one country has the attention of others. For example, we can see how the United States is a big player on all Wikinews, illustrating how the world pays attention to what the US is doing and its presence in global politics.

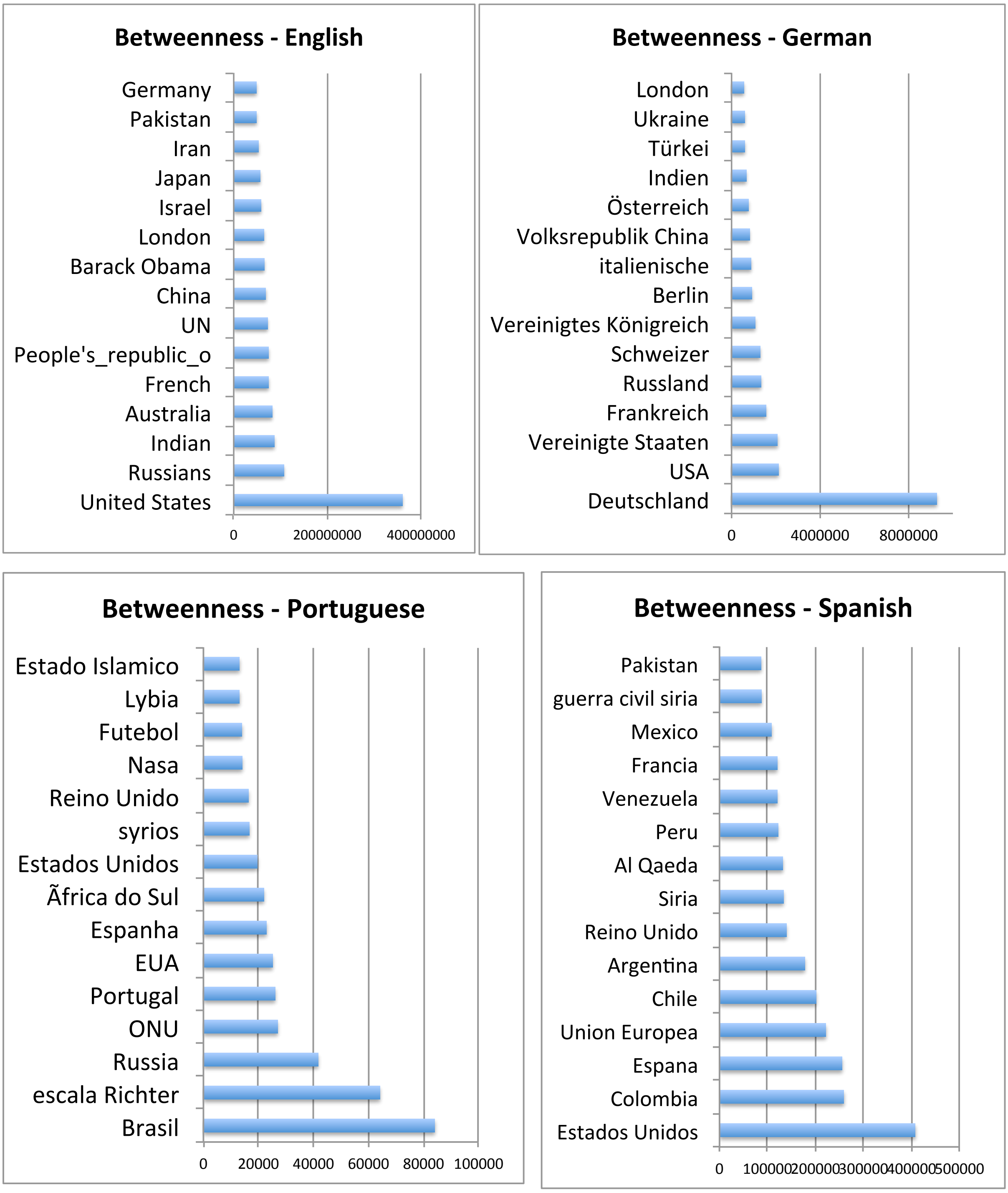

**Figure 10. Most important topics by betweenness in the different language Wikinews**

We used the sentiment, emotionality, and complexity functions of the tool Condor (Brönnimann 2014) to calculate these metrics for all four Wikinews datasets. The overall negativity of sentiment stands out, below the neutrality threshold of 0.5. This confirms the popular belief, that people are much more interested in negative than in positive news. However, we also notice that some Wikinews datasets are more positive than others (Figure 11). For instance, the English language Wikipedia is the most negative while German,

Portuguese and Spanish are significantly most positive. It also seems that Spanish Wikinews uses more complex language than the others, which might be due to the many countries that share the same language. Spanish speakers in Spain might be saying similar things in other words than speakers in Mexico, Chile, or Argentina. The same could be claimed for the English Wikinews, due to United States, United Kingdom, Australia (etc.) editing. However, in this case, the population of the United States is considerably larger than that of the others combined. German Wikinews basically just covers one country, where the language might be used more consistently, leading to less complex language. On the other hand, the Germans seem to be less emotional than the Spanish, Portuguese, or the English, confirming a national stereotype.

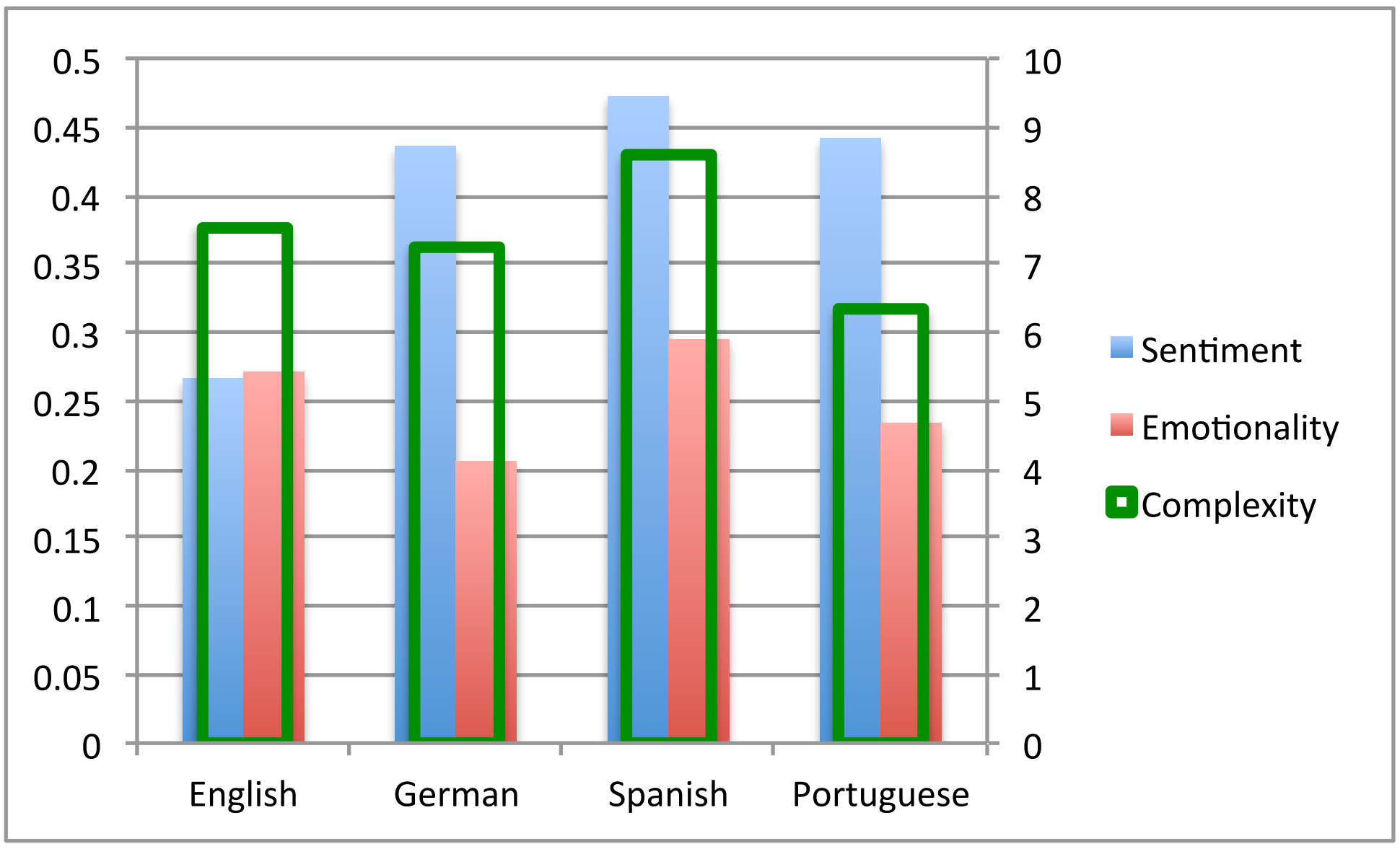

**Figure 11. Sentiment, emotionality, and complexity of the pages linked to WikiNews in the different languages**

## 6 Related work

Charles Murray's list of "Human Accomplishments" (Murray 2003) shows the most innovative people in history. The list is led by Isaac Newton and Gallileo Gallilei. It is based on a compilation of history textbooks, where "accomplished" people have to be listed multiple times in different languages and sources. Murray's list has been combined in the Pantheon with Wikipedia (http://pantheon.media.mit.edu/methods). Pantheon is based on a list created from 11,340 Wikipedia people pages, where a person has to be listed in at least 25 different language Wikipedias. The list has been hand curated to remove obviously non-significant characters. Our analysis approach is purely based on data and not hand-curated, making a claim to higher objectivity. Secondly, to be included into the Pantheon, a person also needs to be in (Murray 2003). The Pantheon seems to be heavily biased towards Greek philosophers, as the top five entries are Aristotle, Plato, Jesus Christ, Socrates, and Alexander the Great.

## 7 Conclusions

The Internet enables researchers to more easily compile rankings of the most important world leaders of all times (Murray 2003, Hidalgo 2014). Our work is unique in that we extract language-specific rankings that allow us to compare the worldview for dozens of different cultures. Probing the historical perspective of many different language-specific Wikipedias gives an X-ray view deep into the historical foundations of cultural understanding of different countries.

http://www.infoplease.com/spot/womenstimeline1.html